\documentclass[aps,prc,twocolumn,superscriptaddress,nofootinbib,showpacs,showkeys,preprintnumbers]{revtex4-1}
\usepackage{graphicx}  
\usepackage{bm}        
\usepackage{amssymb}   
\usepackage{amsmath}   
\usepackage{hyperref}
\usepackage{multirow}
\usepackage{lineno}
\usepackage{comment}
\usepackage{subcaption}
\usepackage{subfiles}

\begin{document}

\title{Elliptic flow of deuterons from simulations with hybrid model}

\author{Tom\'a\v{s} Poledn\'i\v{c}ek} \affiliation{Fakulta jadern\'a a fyzik\'aln\v{e} in\v{z}en\'yrsk\'a, \v{C}esk\'e vysok\'e u\v{c}en\'i technick\'e v Praze, \\  B\v rehov\'a 7, 11519 Praha 1, Czech Republic}
\author{Radka Voz\'abov\'a} \affiliation{Fakulta jadern\'a a fyzik\'aln\v{e} in\v{z}en\'yrsk\'a, \v{C}esk\'e vysok\'e u\v{c}en\'i technick\'e v Praze, \\  B\v rehov\'a 7, 11519 Praha 1, Czech Republic}
\author{Boris Tom\'a\v{s}ik} \affiliation{Fakulta jadern\'a a fyzik\'aln\v{e} in\v{z}en\'yrsk\'a, \v{C}esk\'e vysok\'e u\v{c}en\'i technick\'e v Praze, \\  B\v rehov\'a 7, 11519 Praha 1, Czech Republic}
\affiliation{Univerzita Mateja Bela, Tajovsk\'eho 40, 974~01 Bansk\'a Bystrica, Slovakia}

\begin{abstract}
Elliptic flow of deuterons is measured on  simulated collisions events of Pb+Pb at CMS energy of 2.76 TeV per colliding nucleon pair. 
We use hybrid model that includes hydrodynamics for the deconfined phase and hadron transport as an afterburner. 
For deuterons, two production mechanisms are examined: coalescence, and direct thermal production during hadronisation and subsequent transport  through the hadronic phase. 
Differential elliptic flow of deuterons in different centrality bins is evaluated for  both implemented production models. 
In this approach, coalescence describes the experimental data better than direct deuteron production.
\end{abstract}
\maketitle


\section{Introduction}
\label{s:intro}

Production of deuterons and larger nuclear clusters from the hot matter produced in relativistic heavy-ion collisions is not fully understood, presently. 
The observation that their abundances are rather well described by the grand-canonical statistical model with chemical freeze-out temperature around 150~MeV \cite{Andronic:2017pug,Andronic:2016nof} is quite puzzling, as their binding energy is smaller by two orders of magnitude. 
Moreover, the typical size of a nucleon cluster is larger than the interparticle distance, hence the interaction of the two nucleons is expected to be screened by surrounding hadrons. 
Note that the hot fireball is usually expected not to decompose immediately after the chemical freeze-out but cools down further down during a few following fm/$c$. 
The clusters would have to exist and survive in such an environment. 
This poses a question, how far one can push an interpretation with such a model in the description of  experimental observables other than the multiplicities.

The other candidate model for cluster production is coalescence which binds nucleons together after their decoupling from the system, based on their proximity in phase space \cite{Butler:1963pp,Gutbrod:1976zzr,Sato:1981ez,Gyulassy:1982pe,Csernai:1986qf,Mrowczynski:1987oid,Lyuboshits:1988yc,Scheibl:1998tk,Chen:2012us,Chen:2013oba,Sun:2015jta,Dong:2018cye,Sun:2018jhg,Sun:2018mqq,Sombun:2018yqh,Hillmann:2021zgj}. 

A recent work \cite{Braaten:2024cke} offers a solution to how one can understand the success of statistical model in describing cluster multiplicity in framework of statistical physics. 
Nevertheless, how deuterons are produced microscopically is still an intriguing question. 

Studies of deuteron production in dynamical models have long tradition \cite{Danielewicz:1991dh}. 
More recently, SMASH hadronic transport model was used to study their production and evolution of their multiplicity during the hadronic phase \cite{Oliinychenko:2018ugs,Staudenmaier:2021lrg}, though the studies ignore the very long formation time of the deuterons \cite{Mrowczynski:2020ugu}.
The PHQMD model includes cluster evolution and uses a cluster finding algorithm which identifies them throughout the whole evolution of the collision \cite{Aichelin:2019tnk}.
The model reproduces experimental data on cluster production from nuclear collisions from few hundred MeV up to 200 GeV per colliding NN pair \cite{Glassel:2021rod,Kireyeu:2022qmv,Coci:2023daq,Kireyeu:2023bye}. 
Another kinetic model was also used for the description of light nuclei production at intermediate energies in \cite{Wang:2023gta}.

In this paper we investigate with the help of a hybrid model whether differential elliptic flow can be used as a discriminator  between the two candidate production mechanisms. 
The idea follows from the previous study \cite{Vozabova:2024jum}.

The paper is structured as follows: in the next Section we explain the idea why elliptic flow could be sensitive to the production mechanism of the clusters. 
In Section~\ref{s:model} the setup of our simulations is explained.
We used experimental data by ALICE collaboration and refer to them in Section \ref{s:exp}.
Results of our simulations are shown in Section~\ref{s:res} and the conclusions based on them are drawn in Section \ref{s:conc}.


\section{Motivation}
\label{s:motiv}

In \cite{Vozabova:2024jum}, two of us have investigated the discriminating power of differential elliptic flow of clusters for the acceptance or rejection of one of the two alternative production models. 

Deuterons are femtoscopic probe. 
This is implicitly included in the formalism of coalescence and implemented since many years \cite{Mrowczynski:1987oid,Scheibl:1998tk,Sun:2018mqq,Mrowczynski:1989jd,Mrowczynski:1989jd,Mrowczynski:1992gc,Mrowczynski:1993cx,Blum:2019suo,Kachelriess:2019taq,Kachelriess:2020amp,Kachelriess:2023jis,Kachelriess:2022khq}
In coalescence, the yield of clusters with certain momentum strongly depends on the size and density profile of the region that produces its constituent nucleons. 
The relevant scale to which the size is compared is the width of the cluster wave function. 
The yield in thermal production just scales with the volume of the producing homogeneity region. 
The dependences on the sizes of the homogeneity regions are different in the two mechanisms. 

The elliptic flow is measured as an azimuthal anisotropy of particle production. 
One has to recognise that particles emitted in different azimuthal directions---including clusters---are produced from different homogeneity regions, which may have different sizes. 
Hence, the elliptic flow of deuterons may be a sensitive probe of the production mechanism. 

The results of \cite{Vozabova:2024jum} have supported this conjecture. 
In that study, hadrons were produced from an extended blast-wave model that included second-order transverse flow and shape anisotropies, resonance production and decays, and modified freeze-out hypersurface with the freeze-out time depending on radial coordinate. 
The model then provided basis for two different deuteron production mechanisms. 
In one set of simulations, thermally produced deuterons were produced from the freeze-out hypersurface like the other hadrons. 
In another set of simulations, deuterons were made by coalescence. 
For this, nucleons were taken from the extended blast wave model, including resonance production. 
Then, the coalescence algorithm searched for their proximity in the phase space. 
The results were compared to data from Pb+Pb collisions at $\sqrt{s_{NN}} = 2.76$~TeV \cite{ALICE:2015wav,ALICE:2017nuf,ALICE:2013mez,ALICE:2014wao}. 
Both transverse momentum spectra and differential elliptic flow of deuterons indicated that coalescence is the preferred production mechanism. 

Here, we repeat the study with a state-of-the-art hybrid model. 
Such models, which combine fluid dynamics for the evolution of the deconfined phase with transport simulation for the hadronic phase, currently provide the most realistic description of ultrarelativistic heavy-ion collisions. 
We use a chain with TRENTo 3D for the initial conditions \cite{Ke:2016jrd}, vHLLE \cite{Karpenko:2013wva} for the description of the fluid in the deconfined phase, and SMASH v2.2 \cite{SMASH:2016zqf,Karpenko:2015xea,Schafer:2021csj} as the hadronic afterburner.

Our task is to compare production of deuterons via coalescence and through direct statistical production. 
While the former is straightforward to implement just by using the nucleons produced in the simulation, the latter needs some explanation. 

In a hybrid model, hadrons are produced \emph{statistically} at the particlisation hypersurface which divides fluid-dynamical evolution and transport simulation.
Then they scatter and may be annihilated as well as created in the transport phase. 
The observed particles are those which leave the fireball at the end of the transport phase. 
Their abundances and momentum distributions are finally determined there. 
In view of this picture,  the closest scenario to the statistical production of deuterons is to generate them statistically at the hadronisation hypersurface and propagate them through the hadronic phase in transport simulation, as was done in \cite{Oliinychenko:2018ugs}. 
This construction will be used here and compared to production by coalescence. 


\section{The model}
\label{s:model}

Our simulations use the chain TRENTo+vHLLE+Hadron Sampler+SMASH. 
We studied Pb+Pb collisions at $\sqrt{s_{NN}}=2.76$~TeV, for centralities up to 70\%.

Initial distributions of energy density and entropy density for the fluid dynamical simulations were generated by three-dimensional version of TRENTo \cite{Ke:2016jrd}.
Parameters were chosen \cite{Ke:2019jbh} so that the best agreement for $p_t$ spectra and $v_2(p_t)$ is achieved for protons, pions, and kaons at all centralities, as measured by ALICE collaboration \cite{ALICE:2013mez}. Relative skewness ansatz for the rapidity distribution was assumed \cite{Ke:2019jbh}.  
The values of parameters are summarised in Table \ref{t:params}.
%
\begin{table}[t]
\centering
\caption{The parameters used in the simulation. TRENTo3D initial condition parameters are taken from \cite{Ke:2019jbh}.}
\label{t:params}
\begin{tabular}{ccc}
\hline\hline
\multicolumn{3}{c}{\textbf{TRENTo3D}} \\ \hline
$\sqrt{s_{NN}}$ & collision energy & 2.76 TeV\\ 
$p$ & generalized mean parameter & 0.0 \\
$k$ & multiplicity fluctuation shape & 2.0 \\ 
$w$ & Gaussian nucleon width & 0.59~fm \\ 
$n$ & normalization & 114.0 \\ 
$\mu_0$ & mean rapidity shift & 0.0 \\
$\sigma_0$ & rapidity distribution width & 2.9 \\
$\gamma_{\mathrm{rel}}$ & rapidity distribution skewness & 7.3 \\
$J$ & pseudorapidity Jacobian & 0.75 \\ \hline\hline

\multicolumn{3}{c}{\textbf{vHLLE}} \\ \hline
$\tau_0$ & initial time & 0.6 fm/$c$ \\ 
$\varepsilon_{\mathrm{crit}}$ & critical energy density & 0.32 GeV/fm$^3$ \\ 
$\eta/s$ & shear viscosity & 0.12 \\ 
$\zeta/s$ & bulk viscosity & 0.4 \\ \hline\hline

\multicolumn{3}{c}{\textbf{SMASH}} \\ \hline
$\varepsilon_{\mathrm{crit}}$ & critical energy density & 0.32 GeV/fm$^3$ \\ 
$t_{\mathrm{max}}$ & maximum simulation time & 100 fm/$c$ \\ \hline\hline
\end{tabular}
\end{table}

%

The initial distributions were further evolved with vHLLE \cite{Karpenko:2013wva}. 
The model assumed shear viscosity with $\eta/s=0.12$. 
Bulk viscosity with a maximum around critical temperature and a peak value of $\zeta/s=0.4$ was parametrised according to eq.(43) of \cite{Schenke:2020mbo}, as it is also implemented in vHLLE.  
The fluid-dynamical simulation evolved the bulk matter down to energy density $\varepsilon_{\mathrm {crit}} = 0.32$~GeV/fm$^3$. 
This marked the particlisation hypersurface, at which Hadron Sampler from the vHLLE+SMASH package was used. 

Each hydrodynamic evolution served as initial condition to configurations of hadrons that were further evolved with SMASH \cite{SMASH:2016zqf}, version 2.2. The maximum time in the simulations was set to 100~fm/$c$ after which any simulation is stopped. 

Different centrality bins were simulated by intervals of impact parameter which were mapped to the experimentally measured percentiles of multiplicity distribution \cite{Loizides:2017ack}.
For each deuteron production mechanism (i.e.~coalescence  or direct thermal) and each centrality bin, the target number of simulated events was $1.5\times 10^6$. 
They were set up as 3000 hydrodynamic events, each oversampled by 500 SMASH afterburner runs. 
The simulated centrality bins were set by  the finest binning from the set of the used experimental data: 0-5\%, 5-10\%, 10-20\%, 20-30\%, 30-40\%, 40-50\%, 50-60\%, 60-70\%. 
More peripheral events were not simulated, as clusters are truly rare there. 
For comparison with data organised in wider centrality bins, the corresponding originally simulated bins were merged. 

For the runs where \emph{coalescence} was implemented, hadrons were propagated beyond their point of last scattering or last resonance decay. 
Then, the same coalescence algorithm was run on (anti)protons and (anti)neutrons as in \cite{Vozabova:2024jum}.
First, for each proton-neutron pair their points of production or last scattering (whatever came later) were boosted into the pair rest frame.
The particle produced earlier was boosted to the time when the later particle was produced. 
Relative momentum $\Delta p = |\boldsymbol{p}_1-\boldsymbol{p}_2|$ and distance $\Delta r = |\boldsymbol{r}_1 - \boldsymbol{r}_2| $ were determined in the pair rest frame.
A deuteron was created with probability 3/8 (to select the correct spin-isospin combination) if $\Delta p < \Delta p_{\mathrm{max}}$ and $\Delta r < \Delta r_{\mathrm{max}}$. 
The value of $\Delta r_{\mathrm{max}} = 3.5$~fm was motivated by \cite{Sombun:2018yqh} and $\Delta p_{\mathrm{max}} = 0.310$~GeV/c was set to reproduce the observed multiplicity of deuterons. 
Finally, the created deuteron was boosted back to the frame in which simulation was running, to the position of the parent nucleons' center of mass. 

In runs with \emph{direct thermal production} of deuterons and antideuterons, they were produced together with other hadrons during the particlisation procedure at the hadronisation hypersurface. 
Then, they entered the transport simulation. 
In SMASH, we chose the covariant collision criterion as was done in \cite{Oliinychenko:2018ugs}, even though the treatment technically involves an auxiliary resonance $d'$ with no counterpart in reality. 
Formally, a more consistent treatment is implemented in SMASH with the stochastic collision criterion \cite{Staudenmaier:2021lrg}, but this leads to CPU times prohibitively large in view of the required statistics. 
For consistency, also simulations with coalescence were set with the covariant collision criterion. 
This is important, since a small but measurable difference in $v_2$ was observed between simulations with geometric and stochastic collision criterion \cite{Staudenmaier:2021lrg}. 

It turned out that it is important to evolve the transport simulation for sufficiently long time. 
Cross-section of deuteron-induced reactions, as implemented in SMASH \cite{Oliinychenko:2018ugs}, are of the order of tens of mb, with $\sigma(\sqrt{s})$ for $\pi d$ scattering peaking above 200~mb. 
Hence, they scatter even at quite low density, as the collision criterion would still be fulfilled for distances between pion and deuteron more than 20~fm. 
We have examined the reliability of our results by running for different values of $t_{\mathrm{max}}$, which is the parameter in SMASH that sets the maximum running time after which the evolution is stopped in any case.
By simulating for $t_{\mathrm max} = 50$~fm/$c$, 100~fm/$c$, and 150~fm/$c$ we checked that our results change between 50 and 100~fm/$c$, but stay unchanged between 100 and 150~fm/$c$.
Further on we report the results obtained for 100~fm/$c$. 

Differential elliptic flow was measured on the simulated data with the scalar product method \cite{Selyuzhenkov:2007zi}, in accord with the experimental data to which we compare our results \cite{ALICE:2017nuf}.


\section{Experimental data}
\label{s:exp}

We perform our analysis on data from Pb+Pb collisions at $\sqrt{s_{NN}} = 2.76$~TeV. 
There are most experimental data available from these collisions and we will be able to compare to the previous study \cite{Vozabova:2024jum}.

In order to check the tuning of the model, we use the $p_t$ spectra of protons, pions, and kaons \cite{ALICE:2013mez}, as well as their $v_2$ \cite{ALICE:2014wao}. 
They were measured in centrality bins 0-5\%, 5-10\%, 10-20\%, 20-30\%, 30-40\%, 40-50\%, 50-60\%. 

Transverse momentum spectra of deuterons were published by ALICE collaboration in \cite{ALICE:2015wav}. They used centrality bins 0-10\%, 10-20\%, 20-40\%, 40-60\%, 60-80\%.
Data on elliptic flow of deuterons are taken from \cite{ALICE:2017nuf}. 
They were organised in centrality bins 0-5\%, 5-10\%, 10-20\%, 20-30\%, 30-40\%, 40-50\%.  


\section{Results}
\label{s:res}

\begin{figure*}[ht]
\begin{center}
\includegraphics[width=0.95\textwidth]{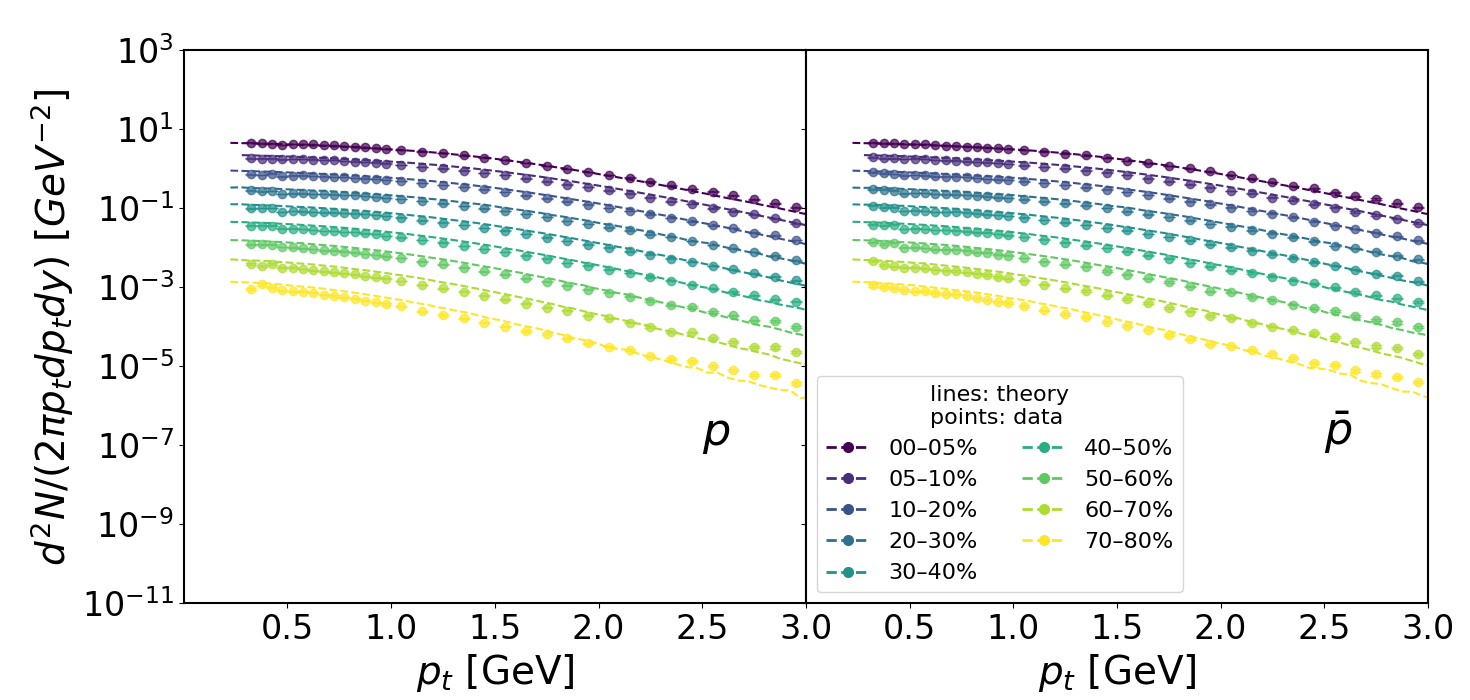}
\end{center}
\caption{Transverse momentum spectra of protons (left) and antiprotons (right). Different colours represent different centralities. Except for the most central data, spectra are divided by subsequent factors of $2^n$. Experimental data measured by the ALICE collaboration \cite{ALICE:2013mez}.}
\label{f:pspec}
\end{figure*}

\begin{figure*}[t]
\begin{center}
\includegraphics[width=0.95\textwidth]{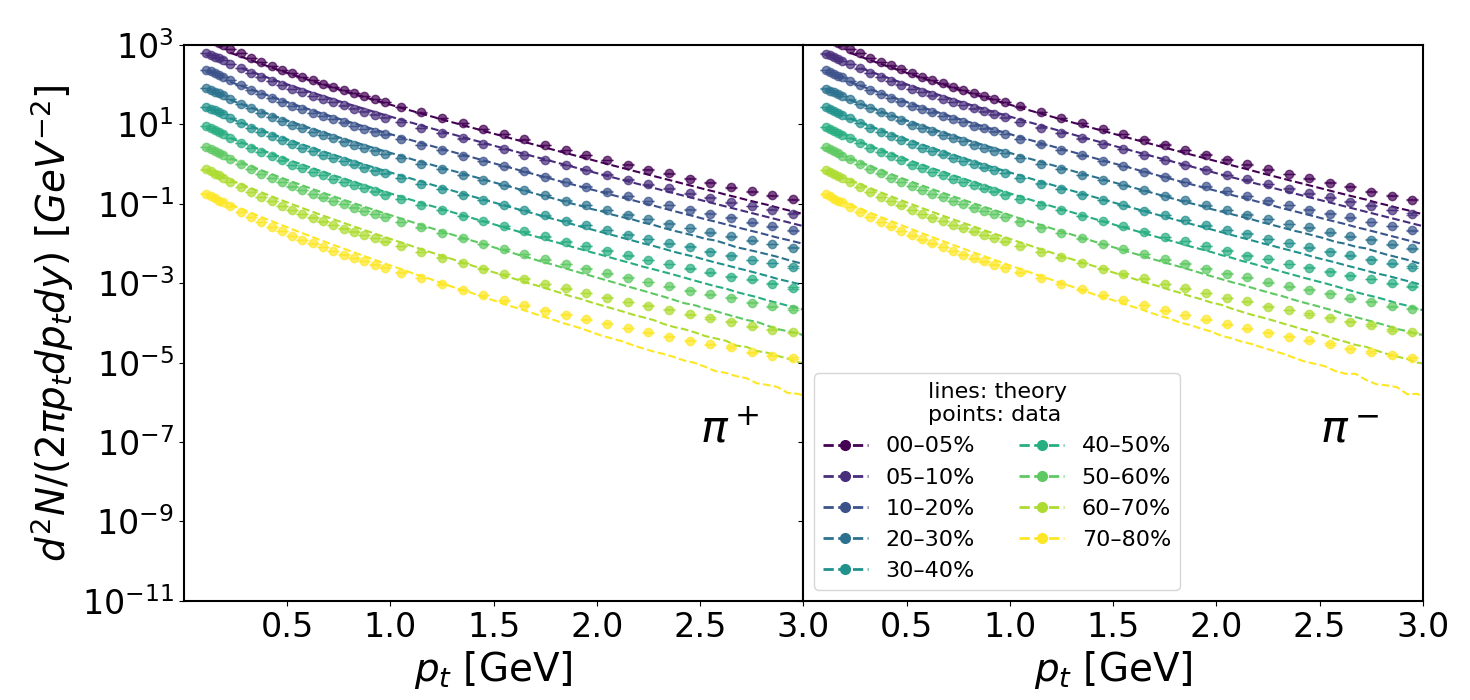}
\end{center}
\caption{Same as Fig.~\ref{f:pspec}, but for positive (left) and negative (right) pions.}
\label{f:pispec}
\end{figure*}


We first check that the tuning reproduces well the transverse momentum spectra and elliptic flow of hadrons. 
In Fig.~\ref{f:pspec} we show our results on $p_t$ spectra of protons and antiprotons plotted together with experimental data by ALICE collaboration \cite{ALICE:2013mez}. 
The $p_t$ spectra of charged pions are shown in Fig.~\ref{f:pispec}.
Note that we also checked the spectra of charged kaons but omit them from showing here for brevity. 
The agreement between simulations and experimental data is reasonably good for $p_t$ below 2~GeV, where hydrodynamics for the bulk production is expected to work.

\begin{figure*}[t]
\begin{center}
\includegraphics[width=0.9\textwidth]{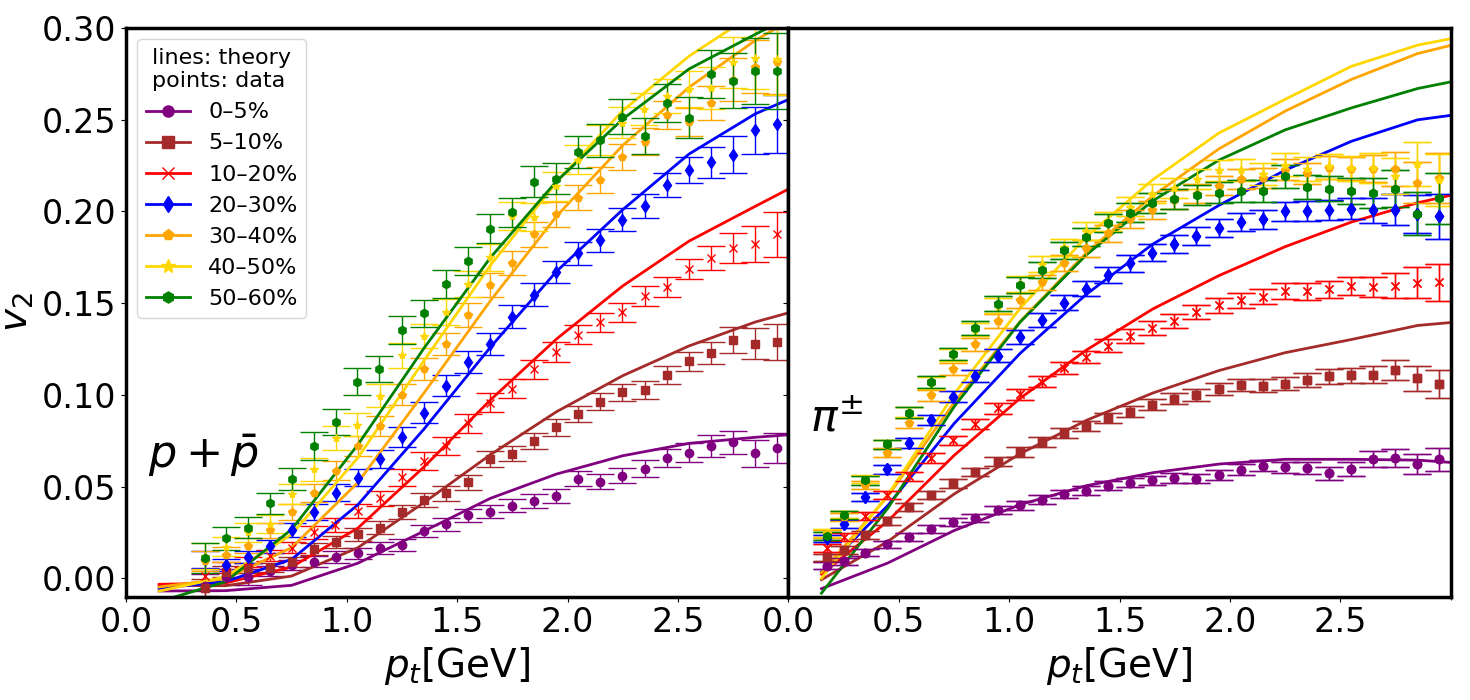}
\end{center}
\caption{Elliptic flow of identified protons and antiprotons (left) and charged pions (right). Different colors represent different centralities. Experimental data measured by the ALICE collaboration \cite{ALICE:2014wao}.}
\label{f:v2had}
\end{figure*}

The elliptic flow of protons and pions is shown in Fig.~\ref{f:v2had}.
The agreement between simulations and data is again reasonable. 
Note that it is extremely hard to reproduce the differential $v_2$ for all centralities and in the whole measured $p_t$ intervals even in much more sophisticated analyses. 
Even some Bayesian analyses of data with the help of some model often fit only integrated $v_2$ at different centralities. 

These results demonstrate that the simulation is tuned appropriately and the predictions for deuteron spectra and elliptic flow can reasonably be compared to data. 

\begin{figure*}[t]
\begin{center}
\includegraphics[width=0.9\textwidth]{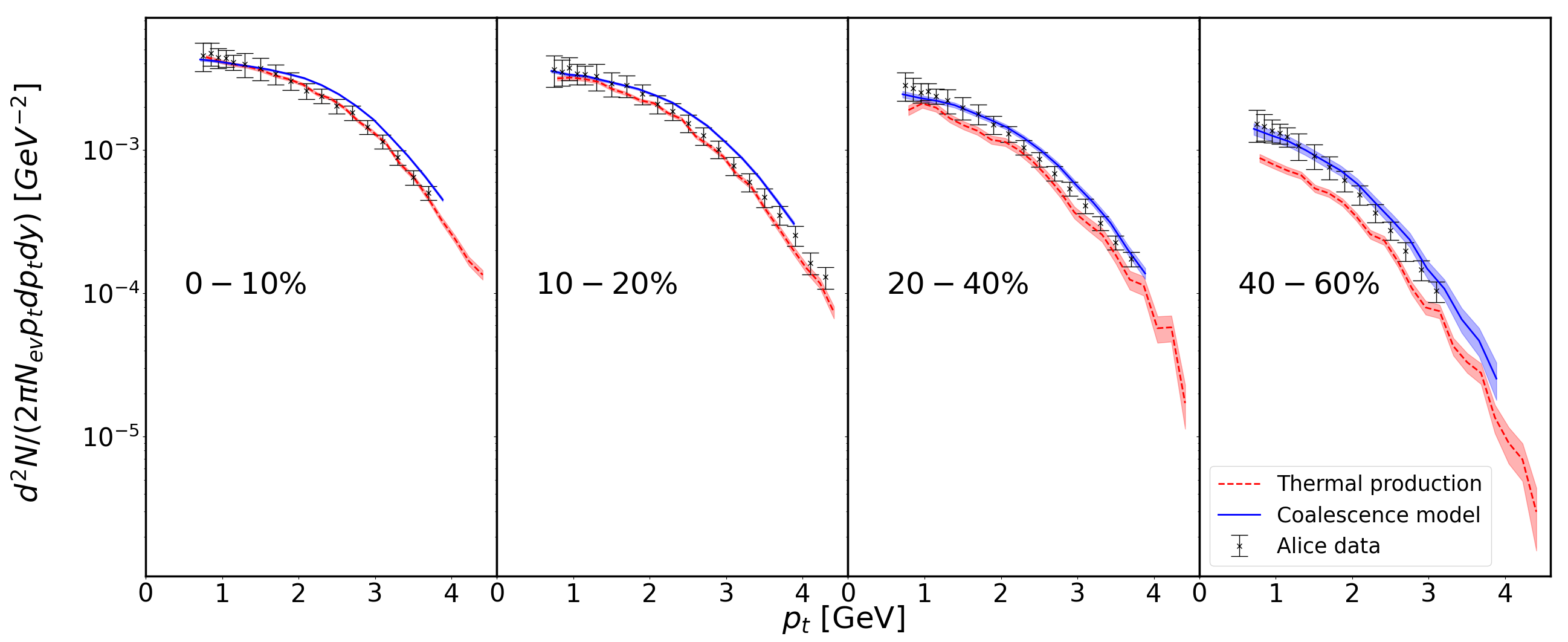}
\end{center}
\caption{Transverse momentum spectra of deuterons and antideuterons. Different panels show results for different centrality bins. Dotted red line: direct thermal production of deuterons; solid blue line: deuterons built by coalescence. Experimental data by ALICE collaboration \cite{ALICE:2015wav}.}
\label{f:dspec}
\end{figure*}

Thus, we proceed with its predictions for deuterons.
Their transverse momentum spectra are shown in Fig.\ref{f:dspec}.
Within the error intervals, both implemented mechanisms agree with the data. 
The direct thermal production slightly underpredicts the $p_t$ spectrum in the most non-central bin here: 40--60\%. 
Nevertheless, the shape of the spectrum follows the measured one. 

The slight underprediction of spectra normalization actually agrees qualitatively with the observation made in \cite{ALICE:2015wav} by  the ALICE collaboration. 
When looking at the deuteron-to-proton ratio as a function of the number of participating nucleons (Fig.~16 in \cite{ALICE:2015wav}), it slightly grows when going from central to non-central collisions. 
Because of the size of the error bars, which are also due to extrapolation of the $p_t$ spectra to full $p_t$ coverage, the dependence is also consistent with a constant. 
Here, in Fig.~\ref{f:dspec} the error bars are much smaller and our theoretical curve lies clearly below the experimental data points. 

\begin{figure}[t]
\begin{center}
\includegraphics[width=0.48\textwidth]{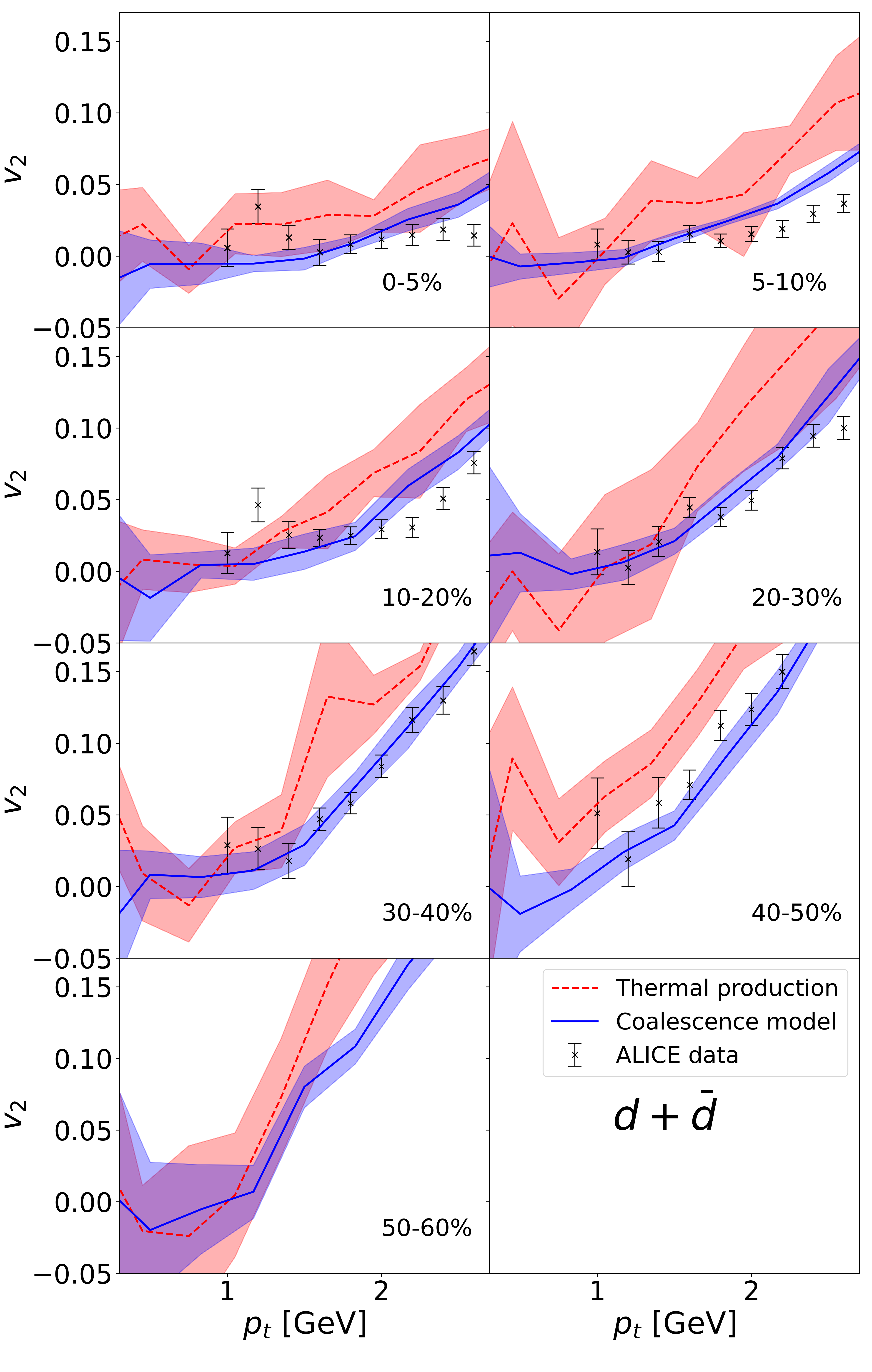}
\end{center}
\caption{Differential elliptic flow of deuterons. Different panels show results for different centrality bins. Red dotted line: predictions  from direct thermal production; blue solid line: predictions from coalescence. Bands around the lines indicate statistical uncertainties. Experimental data by the ALICE collaboration \cite{ALICE:2017nuf}.}
\label{f:dv2}
\end{figure}

Our main goal, however, was to search for the difference between predictions for differential $v_2$. 
The results are shown in Fig.~\ref{f:dv2}. 
The results seem to indicate that the  elliptic flow of deuterons that are thermally produced  and evolved in SMASH is higher than that  from coalescence. 
This feature appears rather systematically in all simulated centrality classes. 
Surprisingly, this is opposite to the observation made in \cite{Vozabova:2024jum}, where coalescence deuterons had higher $v_2$.

When compared to data, direct thermal production overshoots the measured $v_2$, while coalescence which was tuned on proton and pion production reproduces the data well. 
Nevertheless, coalescence slightly overshoots the data in the more central classes of events. 
A closer inspection of proton elliptic flow, plotted in Fig.~\ref{f:v2had}, shows that it is also the case for protons (and even pions) with transverse momenta below 1~GeV/$c$.
Thus, since the deuteron consists of two nucleons it is reasonable to expect that the tension with data reaches up to twice that upper limit. 


\section{Conclusions}
\label{s:conc}

Our simulations with hybrid model including vHLLE hydrodynamics and SMASH as hadron afterburner show qualitatively different results for the differential elliptic flow than the previous analysis of two of us \cite{Vozabova:2024jum}:
elliptic flow here is larger from the thermal production than from coalescence, while it was the other way around in \cite{Vozabova:2024jum}.
We are thus prevented from stating that deuteron elliptic flow is always larger in one mechanism in comparison to the other. 
Differential elliptic flow cannot be the decisive probe to distinguish the two mechanisms in such a simple way. 

It is the difference between what is called thermal model here and what it was in \cite{Vozabova:2024jum} that is the cause for the different results. 
Let us compare the two models more closely. 
To this end, recall that in \cite{Vozabova:2024jum} all hadrons were produced from an extended blast-wave parametrization. 
This included resonances which were let to decay. 
No rescattering between hadrons happened after that. 

In both cases of the thermal model, the assumption is that the deuteron is a pointlike particle. 
This is an important feature because in thermal models with expansion the unique assignment of position leads to the determination  of the momentum through the connection between position and the flow velocity.
(Note that coalescence takes the finite size of the deuteron implicitly into account.)
The difference between \cite{Vozabova:2024jum} and here is that there the clusters are immediately produced in the final state while here they are generated at the beginning of the transport simulation. 
After that they still rescatter and can be destroyed as well as created. 
Note that they have large cross-section for the interaction with pions and the large abundance of the latter ones. 
As we already mentioned, the rescattering goes on until rather late time and contributed to the buildup of the elliptic flow. 
The higher $v_2$ is a natural consequence of this.
From this point of view, the two \emph{thermal models}, as implemented here and in  \cite{Vozabova:2024jum} are rather different. 
Nevertheless, what we use here is the closest implementation of such a model that can be consistently run within the framework of a hybrid model. 

On the other hand, our results consistently show that for both the $p_t$ spectra and the differential elliptic flow of deuterons, coalescence is consistent with data while the direct production of deuterons is not. 
Recall that the simulations were tuned on light hadrons and the description of deuterons is a prediction that follows from that tuning. 

In the used model, differential elliptic flow of direct deuterons grows until very late times, when the production of light hadrons has been finished, already.
Such a feature would ask for deeper study because it comes down to the question of applicability of deuteron transport through the hot hadronic has in SMASH. 

The original argument which motivated this study was based on the fact that coalescence depends non-trivially in the size of the producing homogeneity region, particularly if it is comparable to the width of the cluster wave function. 
An interesting plan might be therefore to try the same analysis with larger clusters. 
There are, however, several caveats in doing this. 

Firstly, scattering of larger clusters, like tritons or $^4$He, in a dense hadronic system is even more problematic concept by itself, as the distance between the nucleons of the cluster is comparable or larger than the typical mean free path. 
Such an objection could be addressed by the PHQMD model which follows the dynamics of nuclear clusters within the medium even if they are unbound \cite{Aichelin:2019tnk,Glassel:2021rod}.

Secondly, larger clusters are heavier and therefore they are suppressed in production. 
For example, $^3$He is suppressed with respect to deuterons by a factor of 300 in Pb+Pb collisions at $\sqrt{s_{NN}}=2.76$~TeV \cite{ALICE:2015wav}. 
In turn, this means that the number of simulated events would have to be increased at least by this factor. 
At the moment, this is prohibitively demanding for us. 

The task thus seems to be to improve the simulated statistics and look for finer probes that may be more robust and better sensitive to the production mechanism. 
Nevertheless, the quantitative comparison of our simulated results to data clearly prefers coalescence over direct thermal production of deuterons.


\section*{Acknowledgments}

This project was supported by the Czech Science Foundation (GA\v{C}R) under Nos 22-25026S and 25-16877S.
BT acknowledges support from VEGA 1/0521/22.
Computational resources were provided by the e-INFRA CZ project (ID:90254),
supported by the Ministry of Education, Youth and Sports of the Czech Republic.
We thank Iurii Karpenko for helpful discussions. 



\end{document}